\documentclass[fleqn,twoside]{article}
\usepackage{espcrc2,graphicx,amssymb,amsbsy,bm,amsmath}

\usepackage{graphicx}
\usepackage[figuresright]{rotating}

\newcommand{\be}{\begin{equation}}
\newcommand{\ee}{\end{equation}}
\newcommand{\bea}{\begin{eqnarray}}
\newcommand{\eea}{\end{eqnarray}}
\newcommand{\vx}{\ensuremath{\vec{x}}}
\newcommand{\vk}{\ensuremath{\vec{k}}}
\newcommand{\bnu}{\ensuremath{\bar{\nu}}}

\title{The Classical and Quantum Inflaton: the Precise Inflationary Potential and 
Quantum Inflaton Decay after WMAP}
\author{D. Boyanovsky
\address[Pitt]{Department of Physics and
Astronomy, University of Pittsburgh, Pittsburgh, Pennsylvania
15260, USA}  \address[obs]{Observatoire de Paris, LERMA. Laboratoire
Associ\'e au CNRS UMR 8112.  
\\61, Avenue de l'Observatoire, 75014 Paris, France.}
\thanks{boyan@pitt.edu} ,
{\underline{H. J. de Vega}}\address[LPT]{LPTHE, Universit\'e
Pierre et Marie Curie (Paris VI) et Denis Diderot (Paris VII),
Laboratoire Associ\'e au CNRS UMR 7589,
Tour 24, 5\`eme. \'etage, 4, Place Jussieu, 75252 Paris, Cedex 05,
France.} \addressmark[obs] \addressmark[Pitt]
\thanks{devega@lpthe.jussieu.fr} ,
N. G. Sanchez\addressmark[obs]\thanks{Norma.Sanchez@obspm.fr}.}
       
\begin{document}

\begin{abstract}
We clarify classical inflaton models by considering them
as {\bf effective} field theories {\` a} la Ginzburg-Landau.
In this approach, the WMAP statement excluding
the pure $ \phi^4 $ potential implies the presence of an inflaton mass term
at the scale $ m \sim 10^{13}$GeV.
Chaotic, new and hybrid inflation models are studied in an unified manner.
In all cases the inflaton potential takes the form
$ V(\phi) =  m^2 \;  M_{Pl}^2 \; v(\frac{\phi}{M_{Pl}}) $, where
all coefficients in the polynomial $ v(\varphi) $ are of order 
$ (m/M_{Pl})^{\! 0} $.
If such potential corresponds to supersymmetry breaking, the
corresponding susy breaking scale is $ \sqrt{m \;M_{Pl}} \sim 10^{16}$GeV
which turns out to coincide with the grand unification (GUT) scale.
The inflaton mass is therefore given by a see-saw formula
$ m \sim M_{GUT}^2/M_{Pl} $. For red tilted spectrum, the potential
which fits the  {\bf best} the present data ($ |1-n_s| \lesssim 0.1 ,
\; r \lesssim 0.1 $) and which  {\bf best} prepares the way for
the forthcoming data is a trinomial polynomial with negative quadratic term
(new inflation). For blue tilted spectrum, hybrid inflation turns to be
the best choice. In both cases, we find an analytic formula relating the
inflaton mass with the ratio $ r $ of tensor to scalar
perturbations and the spectral index $ n_s $ of scalar perturbations:
$ 10^6 \; \frac{m}{M_{Pl}} = 127 \; \sqrt{r|1-n_s|} $ where the numerical
coefficient is fixed by the WMAP amplitude of adiabatic perturbations.
Implications for string theory are discussed.
We then review \emph{quantum} phenomena during inflation
which contribute to relevant observables in the CMB anisotropies
and polarization and we focus on {\it inflaton decay}.
The  deviation from the scale invariant power spectrum measured by a  
small parameter  $\Delta$ turns to be crucial, $\Delta$ regulates the infrared too. 
In slow roll inflation, $\Delta$ is a simple function of the slow roll
parameters. We find that quantum
fluctuations can \emph{self-decay} as a consequence of the
inflationary expansion through processes which are forbidden in
Minkowski space-time.  We compute the \emph{self-decay} of the
inflaton quantum fluctuations during slow roll inflation: for wavelengths deep inside the
Hubble radius the decay is \emph{enhanced} by the emission of
ultrasoft collinear quanta, i.e. \emph{bremsstrahlung radiation of
superhorizon quanta} which becomes the leading decay channel for
physical wavelengths $H\ll k_{ph}(\eta)\ll H/(\eta_V-\epsilon_V)$.
The decay of short wavelength fluctuations hastens as the physical
wave vector approaches the horizon. Superhorizon fluctuations decay
with a power law $\eta^{\Gamma}$ in conformal time where 
$ \Gamma $ is expressed in terms of
 the amplitude of curvature perturbations $
\triangle^2_{\mathcal{R}} $, the scalar spectral index $n_s$, the
 tensor to scalar ratio $r$ and slow roll parameters.
The behavior of the growing mode $ {\eta^{\eta_V-\epsilon_V
+\Gamma}}/{\eta} $ features a new scaling dimension $\Gamma$.
We discuss the  implications of these results for scalar and tensor
perturbations as well as for non-gaussianities in the power
spectrum. The recent WMAP data suggests $\Gamma \gtrsim 3.6 \times
10^{-9} $.
\vspace{1pc}
\end{abstract}

\maketitle

\section{Introduction}

A period of accelerated expansion in the early universe, namely
inflation, is nowadays part of standard cosmology since explains the
homogeneity, isotropy and flatness of the observed Universe
\cite{libros}. At the same time, inflation provides a
mechanism for generating metric fluctuations which seed large scale
structure: during inflation physical scales grow faster than the
Hubble radius but slower than it during both radiation or matter
domination eras, therefore physical wavelengths cross the horizon
(Hubble radius) \emph{twice}. Quantum fluctuations generated during
inflation with wavelengths smaller than the Hubble radius become
classical and are amplified upon first crossing the horizon. As they
re-enter the horizon during the decelerated stage these fluctuations
provide the seed for matter and radiation inhomogeneities which
generate structure upon gravitational collapse. Most of the
inflationary models predict fairly generic features: a gaussian,
nearly scale invariant spectrum of adiabatic  scalar and tensor
primordial perturbations (gravitational waves). These generic
predictions are in spectacular agreement with Cosmic Microwave
Background (CMB) observations. Gaussian \cite{komatsu} and adiabatic
nearly scale invariant primordial fluctuations \cite{spergel}
provide an excellent fit to the WMAP data as well as to a variety of
large scale structure observations. Perhaps the most striking
confirmation of inflation as the mechanism for generating
\emph{superhorizon} (`acausal') perturbations is the anticorrelation
peak in the  temperature-polarization (TE) angular power spectra at
$l\sim 150$ corresponding to superhorizon scales
\cite{kogut,peiris}. The anticorrelation between the E-mode (parity
even) polarization fluctuation and the temperature fluctuation is a
distinctive feature of superhorizon adiabatic fluctuations
\cite{sperzalda}: the (peculiar) velocity gradient generates a
quadrupole temperature anisotropy field around electrons which in
turn generates an E-polarization mode. By continuity, the gradient
of the peculiar velocity field is related to the time derivative of
the density (temperature) fluctuations, hence  the peculiar velocity
and the initial (adiabatic) contribution to the (acoustic)
oscillations of the photon baryon fluid are out of phase by $\pi/2$
\cite{sperzalda}. Thus, the product of these two terms gives an
anticorrelation peak at $k \, c_s \; \eta_{dec} =3 \, \pi/4$ which
corresponds to superhorizon wavelengths since the size of the
horizon is $\sqrt{3}$ larger than the size of the sound horizon. The
WMAP (TE) data \cite{kogut,peiris} clearly displays this (TE)
anticorrelation peak at $l \sim 150$ providing perhaps one of the
most striking confirmations of adiabatic superhorizon fluctuations
as predicted by inflation. While the robust predictions of a generic
inflationary model provide an excellent fit to the WMAP data, 
it is still necessary to decide between a host of possible models\cite{clar}.
In addition, 
potential experimental deviations from the most generic features are
the focus of intense study as the low angular momentum depletion. 
With the ever increasing precision of
CMB measurements, it is conceivable that forthcoming observations
will allow a substantial progress in singling out inflationary 
models\cite{clar}. 
Relevant discriminants between models are: non-gaussianity, deviations from
constant scaling exponents (running spectral index) either for scalar 
and/or tensor 
perturbations, an isocurvature component of primordial fluctuations, etc. 
Quantum effects associated with interactions can
potentially lead to non-gaussian
correlations\cite{nos},\cite{allen}-\cite{bartolo}. Therefore the detection of
deviations from constant scaling exponents (as hinted in the WMAP data) or small
non-gaussianities in the temperature correlations imply potentially
interesting quantum phenomena during the inflationary stage that were
imprinted on superhorizon scales.

The inflaton is usually studied as a homogeneous  classical scalar
field\cite{libros}. However, important aspects of the
dynamics require a full quantum treatment for their consistent
description. The quantum dynamics of the inflaton is
systematically treated within a non-perturbative framework and
consequences on the CMB anisotropy spectrum were analyzed in
ref.\cite{cosmo}.

\section{Classical Inflation as an Effective Field Theory: the 
Best Inflationary Potential from the WMAP Data}

The classical dynamics of the inflaton (a massive scalar field) 
coupled to a cosmological background clearly shows that inflationary
behaviour is an {\bf attractor}\cite{bgzk}. This is a generic and robust
feature of inflation. 
The robust predictions of inflation (value of the  entropy of the universe, 
solution of the flatness problem, a small amplitude almost scale invariant 
power spectrum of adiabatic Gaussian density fluctuations explaining the CMB
anisotropies) which are common to many 
available inflationary scenarios, show the predictive power of the 
inflationary paradigm. Whatever the microscopic model for the early universe 
(GUT theory, string theory) would be, it should include inflation with the 
generic features we know today. 

Inflationary dynamics is typically studied by treating  the
inflaton as a homogeneous  classical scalar
field\cite{libros} whose evolution is determined
by a classical equation of motion, while the inflaton 
quantum fluctuations (around the classical value and in the
Gaussian approximation) provide the seeds for the scalar
density perturbations of the metric. In quantum field theory,
this classical inflaton corresponds to the expectation value 
of a quantum field operator in a translational invariant state.
Important aspects of the inflationary dynamics, as resonant
particle production and the nonlinear back-reaction that it generates,
require a full quantum treatment of the inflaton for their consistent
description. The quantum dynamics of the inflaton 
in a non-perturbative framework and its consequences on the CMB anisotropy 
spectrum were treated in refs.\cite{cosmo}. Particle decay in a de Sitter 
background has been studied in refs. \cite{prem,holman} and during slow roll 
inflation in ref.\cite{nos} together with its implication for the
decay of the density fluctuations.

Inflation as known today should be considered as an {\bf effective theory},
that is, it is not a fundamental theory but a theory of a
condensate (the inflaton field) which follows from a more fundamental one 
(the GUT model, string theory). The inflaton field $ \phi $ may {\bf not}
correspond to any real particle (even unstable) but is just an {\bf effective}
description while the microscopic description should come from the GUT model. 
At present, there is no derivation of the inflaton model from more 
microscopic theories, either GUT models or string theories. However, 
the relation of inflation to these microscopic theories is akin to the relation
between the effective Ginzburg-Landau theory of superconductivity
and the microscopic BCS theory, or the relation between the $O(4)$
sigma model of low lying mesons and quantum chromodynamics (QCD).

We provide in ref. \cite{clar} a clear understanding of inflation
and the inflaton potential from effective field theory and the 
WMAP data. This clearly places inflation within the perspective and
understanding of
effective theories in particle physics. In addition, it sets up a clean way to 
directly confront the inflationary predictions with the forthcoming CMB data 
and select a definitive model.

The following inflaton potential or alternatively the hybrid inflation model 
are rich enough to describe the physics
of inflation and accurately reproduce the available data 
\cite{komatsu,spergel,kogut,peiris}:
\be\label{potint}
V(\phi) =  |m^2| \;  M_{Pl}^2 \left[ v_0 \pm \frac12 \; \varphi^2 + 
\frac23 \; \gamma \; 
\varphi^3 + \frac1{32} \; \kappa  \; \varphi^4 \right] \; .
\ee
Here $ \varphi \equiv \frac{\phi}{M_{Pl}} \; , |m| \sim 10^{13}$GeV, 
the dimensionless parameters 
$ \gamma $ and $ \kappa $ are of order 
$ \left(\frac{m}{M_{Pl}}\right)^{\! 0} $, 
and $ v_0 $ is  such that $ V(\phi) $ and $ V'(\phi) $ vanish at the absolute 
minimum of $ V(\phi) $.
This ensures that inflation ends after a finite time
with a finite number of efolds. $ \kappa $ must be positive to ensure stability
while $ \gamma $ and the mass term $ \varphi^2 $ can have either sign. 
$ \gamma $ describes how asymmetric is the potential while  $ \kappa $
determines how steep it is. After factoring out the scales $m$ and $M_P$
there remains the small quantity $1/N_{efolds}\sim\mathcal{O}(10^{-2})$ which
determines the departure from scale invariance as well as the scalar
to tensor ratio in this effective description\cite{clar}.
($N_{efolds}$ being the number of efolds from the first horizon 
crossing to the end of inflation).

The potential eq.(\ref{potint}) cover a wide class of inflationary scenarios:
small field scenarios (new inflation) for spontaneously broken symmetric
potentials (negative mass square), as well as large field scenarios
(chaotic inflation) for unbroken symmetric potentials (positive mass square).
Coupling the inflaton to another scalar field yields the hybrid type scenarios.
Renormalizability restricts the degree of the potential eq.(\ref{potint})
to four. Indeed, in the context of effective theories potentials
of any degree may be considered but a quartic potential is rich enough 
to describe the full physics and to reproduce accurately the WMAP data.

In the context of an effective theory or Ginzburg-Landau model it is {\bf 
highly unnatural} to drop the quadratic term $ \varphi^2 $. 
This is to  exactly choose the  {\bf critical} point of the model
$ m^2 = 0 $. In fact, the recent WMAP \cite{komatsu,spergel,kogut,peiris} 
statement disfavouring the monomial  $ \varphi^4 $ potential 
precisely supports a 
generic polynomial inflaton potential as in eq.(\ref{potint}).
Excluding the quadratic mass term in the potential 
$ V(\phi) $ implies to fine tune the mass to zero which is 
only justified at isolated (critical) points. 
Therefore the pure quartic potential $ \varphi^4
$ is physically an unnatural  choice implying  fine tuning  to zero
the coefficient of $ \varphi^2 $.

We obtain analytic and unifying expressions for chaotic and new inflation
for the relevant observables\cite{clar}:
the amplitude for scalar fluctuations $ |{\delta}_{k\;ad}^{(S)}|^2 $, 
spectral index  $ n_s $ and ratio $ r $ of tensor to scalar
perturbations as well as for hybrid inflation and plot them for the three
scenarios. Particularly interesting are the plots of 
$ n_s $ vs. $ r $\cite{clar}.

We express the ratio of the inflaton mass and the Planck mass 
$ x \equiv 10^6 \; \frac{m}{M_{Pl}} $ in terms
of the amplitude of adiabatic perturbations and the parameters in the 
potential. Furthermore,  we can express $ x $
in terms of observable quantities as $ r $ and $ n_s $. 
We find for new inflation when both $ r $ and $ | n_s - 1 | $ are small,
\bea \label{xint}
10^6 \; \frac{m}{M_{Pl}} = x = 5 \, \pi \, \sqrt{3} \; 10^5 \, 
|{\delta}_{k\;ad}^{(S)}| \; \sqrt{r(1-n_s)} = &&\cr \cr
=127 \; \sqrt{r(1-n_s)} \pm 6 \% \; ,  &&
\eea
where the $ \pm 6 \% $ correspond to the error bars in the 
amplitude of adiabatic perturbations\cite{komatsu,spergel,kogut,peiris}.
We analyze in ref. \cite{clar} how the mass ratio $ \frac{m}{M_{Pl}} $ varies with $ n_s $ 
and $ r $. We find a {\bf limiting} value  $ x_0 \equiv 10^5  \; 
\frac{m_0}{M_{Pl}} \simeq 1 $ for the inflaton mass 
such that $ m_0 \simeq  10^{-5} \; M_{Pl} $ is
a {\bf minimal} inflaton mass for chaotic inflation, and a  {\bf maximal}
mass for new inflation in order to keep $ n_s $ and $ r $ within the WMAP data.

New inflation arises for broken symmetric potentials (the minus sign in front
of the $ \varphi^2 $ term) while chaotic inflation appears both for unbroken
and broken symmetric potentials. 
For broken symmetry, we find that analytic continuation
connects the observables for chaotic and new inflation:
the observables 
are {\bf two-valued} functions of $ y \equiv \kappa \, N_{efolds} $. 
%($N_{efolds}$ being the number of efolds from the first horizon 
%crossing to the end of inflation).
One branch corresponds to new inflation and the other 
branch to chaotic inflation. $n_s, \; r $ and 
$ |{\delta}_{k\;ad}^{(S)}|^2 $
for chaotic inflation are connected by analytic continuation to
the same quantities for new inflation. The branch point
where the two scenarios connect 
corresponds to the monomial $ +\frac12 \, 
\varphi^2 $ potential ($\kappa=\gamma=0$). 

The potential which {\bf best fits} the present data for a red tilted 
spectrum ($ n_s < 1 $)
and which {\bf best prepares} the way to the expected data (a small 
$ r \lesssim 0.1 $) is given by the trinomial potential eq.(\ref{potint}) with
a negative  $ \varphi^2 $ term, that is {\bf new} inflation. 

In new inflation we have the upper bound 
$$ 
r \leq \frac{8}{N_{efolds}} \simeq  0.16 \quad  .
$$
This upper bound is attained by the quadratic monomial potential. 
On the contrary, in chaotic inflation for both signs of the $ \varphi^2 $ term,
$ r $ is bounded as
\be  \label{cotcao}
0.16 \simeq \frac8{N_{efolds}} < r < \frac{16}{N_{efolds}}  \simeq 0.32\quad ,
\ee
This bound holds for all values of the cubic coupling $ \gamma $ which describes
the asymmetry of the potential. 
The lower and upper bounds for $ r $ are saturated by the quadratic and quartic
monomials, respectively.

If an upper bound $ r \leq  0.16 $ turns out to be measured
eq.(\ref{cotcao}) implies that chaotic inflation is {\bf excluded}.

For chaotic and new inflation, we find the following properties\cite{clar}:
\begin{itemize}
\item{ $n_s$ is bounded as
$$ 
n_s \leq 1 - \frac2{N_{efolds}} \simeq 0.96 \qquad  \mbox{chaotic~inflation,}
$$
$$ 
n_s \leq 1 - \frac{1.558005\ldots}{N_{efolds}} \simeq 0.9688 \qquad  
\mbox{new~inflation} \; .
$$ 
The value at the bound for chaotic inflation corresponds to the quadratic 
monomial potential.}
\item{ $n_s$ decreases with the steepness $ \kappa $ 
for fixed asymmetry $ h \equiv\gamma \; \sqrt{\frac{8}{\kappa}} < 0 $ 
and  $n_s$ grows with the asymmetry $ |h| $ for fixed steepness $ \kappa $.}
\end{itemize}

%For chaotic inflation $ r $  grows with the steepness $ \kappa $ for fixed
%asymmetry $ h < 0 $ and decreases with the asymmetry $ |h| $ for fixed
%steepness $ \kappa $. Also, in chaotic inflation $ r $ decreases with $ n_s $.
%For new inflation  $ r $ does the opposite: {\bf it decreases} with
%the steepness $ \kappa $ for fixed asymmetry $ h < 0 $ while it grows 
%with the asymmetry $ |h| $ for fixed  steepness 
%$ \kappa $. Also, in new inflation $ r $ grows with $ n_s $.
For the general trinomial potential eq.(\ref{potint}),
$ r $ decreases with $ n_s $ in chaotic inflation 
while, in new inflation, $ r $ grows with $ n_s $.
%In addition, $ r $ {\bf decreases} for 
%increasing  asymmetry $ |h| $ at a fixed $ n_s $ in new inflation 
%(with $ h<0 $). As a consequence, t
The trinomial 
potential  eq. (\ref{potint}) can yield {\bf very small $r$} for red tilt with
{\bf $ n_s < 1 $ and near unit} for new inflation.  

Hybrid inflation always gives a blue tilted spectrum $ n_s > 1 $ in the
$\Lambda$-dominated regime, 
allowing $ n_s - 1 $ and $ r $ to be small. Interestingly enough, 
we obtain for hybrid inflation a formula for the mass ratio $ x $
with a similar structure to eq.(\ref{xint}) for new inflation:
$$
x =  10^6 \; \frac{m}{M_{Pl}} = 127 \, 
\sqrt{r \, \left(n_s -1 + \frac38 \; r\right) } \quad .
$$
As shown in ref.\cite{clar}, $ \frac{m}{M_{Pl}} $ 
{\bf decreases} when $ r $  {\bf and} $ n_s - 1 $ both approach zero.
We relate the cosmological constant in the hybrid inflation Lagrangian
with the ratio $r$ as 
$$
\frac{\Lambda_0}{M^4_{Pl}} = 0.329 \times 10^{-7} \; r \; ,
$$
and we find that $ (n_s - 1) $ gives
an {\bf upper bound} on the cosmological constant:
$$
  \frac{\Lambda_0}{m^2 \;M_{Pl}^2} < \frac2{n_s-1} \; .
$$
In summary, for small  $ r \lesssim 0.1 $ and $ n_s $ near unit,
{\bf new} inflation from the trinomial potential eq.(\ref{potint})
and {\bf hybrid} inflation emerge as the {\bf best} inflation candidates.
Whether $n_s$ turns to be larger or smaller than one will 
choose hybrid inflation or new inflation, respectively. 
In any case $|n_s -1|$ turns to be of order $ 1/N_{efolds} $.
This can be understood intuitively as follows: the geometry of the universe
is scale invariant during de Sitter stage since the metric takes in conformal 
time the form 
$$
ds^2 = \frac1{(H \; \eta)^2}\left[ (d \eta)^2 - (d \vec x)^2 \right] \; .
$$
Therefore, the primordial power generated is scale invariant except
for the fact that inflation is not eternal and lasts for $N_{efolds}$.
Hence, the  primordial spectrum is scale invariant up to $ 1/N_{efolds} $ corrections.
Also, the ratio $ r $ turns to be of order  $ 1/N_{efolds} $ (chaotic and new inflation)
or  $ 1/N_{efolds}^2 $ (hybrid inflation). 

Generally speaking, the amplitude of 
adiabatic perturbations is given in order of magnitude by\cite{clar}
$$
|{\delta}_{k\;ad}^{(S)}| \sim N_{efolds} \; \frac{|m|}{M_{Pl}} \; .
$$

\section{Implications for Supersymmetry and String Theory}

In order to reproduce the CMB data, the inflationary potentials 
in the slow roll scenarios must have the structure
\be \label{vef}
V(\phi) =  M^4 \; v\!\left(\frac{\phi}{M_{Pl}}\right) \; ,
\ee
where $ v(0) = v'(0) = 0 $  and all higher derivatives at the origin are of the
{\bf order one}. The inflaton mass is therefore given by a see-saw-like
formula
\be \label{mint}
m \simeq \frac{M^2}{M_{Pl}} \; .
\ee
As stated above, the WMAP data imply $ m \sim 10^{13}$GeV, Eq. (\ref{mint})
implies that $M$ is {\bf precisely} at the grand unification 
scale $M \sim 10^{16}$GeV \cite{libros}. 

Grand unification proposes that at some energy scale all three
couplings (electromagnetic, weak and strong) should merge into one.
In this case, such grand unified scale turns out to be  $ E
\sim 10^{16}$GeV \cite{gut,sw}. 

Three strong independent indications
of this scale are available nowadays: 1) the convergence of the 
running electromagnetic, weak and strong couplings, 2) the large mass scale
to explain the neutrino masses via the see-saw mechanism and 3) the
scale $M$ in the above inflaton potential. Also, notice that eq.(\ref{mint}) 
has the structure of the moduli potential coming from supersymmetry breaking.
Therefore, the supersymmetry breaking scale would be at the GUT scale too.

The running of the couplings with the energy (or the length) is governed
 by the renormalization group.
For the standard model of electromagnetic, weak and strong interactions,
the renormalization group yields 
that the three couplings get unified approximately at $ \sim 10^{16}$GeV. 
A better convergence is obtained in supersymmetric extensions of the
standard model \cite{gut,sw}.

Neutrino oscillations and neutrino masses are currently explained in the 
see-saw mechanism as follows\cite{ita},
$$
\Delta m_{\nu} \sim \frac{M^2_{Fermi}}{M} 
$$
where $ M_{Fermi} \sim 250$ GeV is the Fermi mass scale, 
$M \gg  M_{Fermi} $ is a large energy scale and $ \Delta m_{\nu} $ 
is the difference between the neutrino masses for different flavors. 
The observed values for $ \Delta m_{\nu} \sim 0.009 - 0.05 $ eV naturally 
call for a mass scale $ M \sim 10^{15-16}$ GeV close to the GUT 
scale\cite{ita}. 

Eq.(\ref{vef}) for the inflaton potential resembles the
moduli potential coming from supersymmetry breaking,
\be\label{susy}
V_{susy}(\phi) =  m_{susy}^4 \; v\!\left(\frac{\phi}{M_{Pl}}\right) \; ,
\ee
where $ m_{susy} $ stands for the supersymmetry breaking scale. 
Potentials with such form were used in the inflationary 
context in refs.\cite{susy}. In our context, eq.(\ref{susy}) implies that 
$  m_{susy} \sim 10^{16}$ Gev. That is, the susy breaking scale
$ m_{susy} $ turns out to be at the GUT scale $ m_{susy} \sim M_{GUT} $.

We see that the mass scale of the inflaton $ m \sim 10^{13}$GeV
can be related with $ M_{GUT} $ by a see-saw style relation eq.(\ref{mint}).

As discussed in sec. 2 the inflaton describes a condensate in a GUT
theory in which it may describe fermion-antifermion pairs.
Current identifications in the 
literature of such condensate field with a given
fundamental field in a SUSY or SUGRA model  have so far no solid basis.
Moreover, the number of supersymmetric models is so large
that there is practically no way to predict which is {\bf the} correct model 
\cite{ramo}.

In order to generate inflation in string theory, one needs first to generate
a mass scale like $ m \sim 10^{13}$GeV and $ M_{GUT} $ related by 
eq.(\ref{mint}).  Without the presence
of the mass scales $m$ and $ M_{GUT} $ [related through
eq.(\ref{mint})],  there is {\bf no} hope in string theory 
to get a correct inflationary cosmology
describing the {\bf observed} CMB fluctuations\cite{noscu}.
Such scale {\bf is not} present in the string action, neither in the action 
of the effective background fields (dilaton, graviton, antisymmetric tensor) 
which are massless. Such scale should be generated 
dynamically perhaps from the string vacuum(ua) but this is still
an open problem far from being solved\cite{noscu}.
Actually, the very same problem hinders the derivation
of a GUT theory and the generation of the GUT scale from string theory.

Since no microscopic derivation of an inflationary model from a GUT is
available so far, it would seem too ambitious at this stage 
to look for a microscopic 
derivation of inflation from string theory. The derivation of
an inflationary cosmology reproducing the observed 
CMB fluctuations is at present too far away in string theory. 
However, an {\bf effective}
description of inflation in  string theory (string matter plus massless
backgrounds) could be at reach\cite{noscu}. 

\section{The inflaton as a quantum field: observable consequences through the 
CMB fluctuations}

We review now \emph{quantum} phenomena during inflation
which contribute to relevant observables in the CMB anisotropies
and polarization.  In particular, we focus on {\it inflaton decay}
during inflation as a potential source of quantum phenomena
contributing to deviations from  scale invariance in the
primordial power spectrum and/or to non-gaussian features. If the
inflaton couples to other particles, then its quantum fluctuations
which seed scalar density perturbations also couple to these other
fields. Consequently, the \emph{decay} of the amplitude of the
\emph{quantum fluctuations} of the inflaton may lead to a
modification of the power spectrum of density perturbations. The
same coupling that is responsible for the decay of the  inflaton
quantum fluctuations can be also the source of non-gaussian correlations.

Particle decay is a distinct feature of interacting quantum field
theories and is necessarily an important part of the inflationary
paradigm: the decay of the inflaton into lighter particles
\emph{after inflation} may yield to the radiation dominated stage.

\medskip

In ref.\cite{nos} we introduced and
implemented a systematic program to study the relaxational
dynamics and particle decay in the case of a rapidly expanding
inflationary stage. Whereby rapid expansion refers to  the Hubble
parameter during inflation being  much larger than the mass of the
particles. In the case of the inflaton, this is the situation of
relevance for slow-roll inflation and a necessary (although not
sufficient) condition for an almost scale invariant power spectrum
of scalar fluctuations \cite{libros}. 

The Minkowski space-time computation of the decay rate is not
suitable for  a stage of rapid expansion (as quantified above):
the rapid expansion of the Universe and the manifest lack of a
global time-like Killing vector allow processes that would be
forbidden by energy conservation in Minkowski space-time. As
emphasized in \cite{nos,woodard}, the lack of energy
conservation in a rapidly expanding cosmology requires a different
approach to study particle decay. The correct decay law follows
from the  relaxation in time of the expectation value of the field
out of equilibrium. The  relaxation of the non-equilibrium
expectation value of the field is computed in ref.\cite{nos}
using the dynamical renormalization group (DRG) which allows to
extract the decay \emph{law} directly from the real time equations
of motion. The reliability and predictive power of the DRG has
been tested for a wide range of physical situations including hot
and dense plasmas in and out of equilibrium \cite{DRG}.

\medskip

 {\bf The goals of this work:}
We compute the particle decay
of quantum fields minimally coupled to gravity with masses $M$ much
smaller than the Hubble parameter, which is the
 relevant case  for slow roll inflation\cite{nos}. This entails a much stronger
infrared behavior than for massless particles conformally coupled to 
gravity\cite{nos}. The emergence of infrared divergences in
quantum processes with gravitons during de Sitter inflation has been
the focus of a thorough study \cite{IRcosmo,dolgov}. As we will see
below, similar strong infrared behavior enters in the decay
 of minimally coupled particles with masses $M$ much smaller than the
 Hubble parameter $H$. When $M<<H$ there is a small parameter
 $\Delta \sim M^2/H^2$ which regulates the infrared behavior in de
 Sitter inflation. 

 \medskip

We began by studying the general case of a cubic interaction of
scalar particles minimally coupled to gravity,  allowing the decay
of one field  into two others during de Sitter inflation\cite{nos}. The
masses of all particles are much smaller than the Hubble constant,
which leads to a  strong infrared behavior  in the self-energy
loops. We introduced an expansion in terms of the small parameter
$\Delta$ which regulates the  infrared and which in the case of de
Sitter inflation is determined by the ratio of the mass squared of the
particle in the loop to the the Hubble constant. Long-time
divergences associated with secular terms in the solutions of the
equations of motion are systematically resummed by implementing
the DRG introduced in refs.\cite{nos,DRG} and lead to the
decay law. 

We then applied these general results to the case of
quasi-de Sitter slow roll inflation\cite{nos}. We showed that in this case a
similar small parameter $\Delta$ emerges which is a simple
function of the slow-roll parameters, and which regulates the infrared
behavior {\it even} for massless particles (gravitons). We studied the decay of
superhorizon fluctuations as well as of fluctuations with
wavelengths deep inside the horizon. A rather striking aspect is
that a particle {\bf can decay} into
 \emph{itself} precisely as a consequence of the lack of energy
 conservation in a rapidly expanding cosmology. We then focus on
 studying the decay of the  inflaton quantum fluctuations into
 their \emph{own quanta}, namely the {\it self-decay} of the inflaton
 fluctuations, discussing the potential implications on the power
 spectra of primordial perturbations and to non-gaussianity.

 \vspace{2mm}

 {\bf Brief summary of results:}
\begin{itemize}
\item{In the case of de Sitter inflation for particles with mass
$M\ll H$, a small parameter $\Delta\sim M^2/H^2$ regulates the
infrared. We introduce an expansion in this small parameter $\Delta$
akin to the $\varepsilon$  expansion in dimensionally regularized
critical theories. We obtain the decay laws in a $\Delta$ expansion
after implementing the DRG resummation.}

\item{Minimally coupled particles decay \emph{faster} than those
conformally coupled to gravity due to the strong infrared behavior
both for superhorizon modes as well as for modes with wavelengths
well inside the Hubble radius.}

\item{The decay of short wavelength modes, those inside the
horizon during inflation, is \emph{enhanced} by soft collinear
\emph{bremsstrahlung radiation of superhorizon quanta} which
becomes the dominant decay channel when the physical wave vector
obeys, 
\be 
k_{ph}(\eta) \equiv \frac{k}{a(\eta)} \lesssim \frac{H}{\eta_V-\epsilon_V}\; ,
\ee 
where $\eta_V,\epsilon_V$ are the standard  slow roll
parameters. }

\item{An  expanding cosmology allows processes that are forbidden
in Minkowski space-time by energy
conservation\cite{nos,woodard}: in particular, for masses
$\ll H$, \emph{kinematic thresholds} are absent allowing a
particle to decay into \emph{itself}. Namely, the
\emph{self-decay} of quantum fluctuations is a feature of an
interacting theory in a rapidly expanding cosmology. A
self-coupling of the inflaton leads to the self-decay of its
quantum fluctuations both for modes inside as well as
\emph{outside} the Hubble radius. }

\item{The results obtained for de Sitter expansion directly apply to the
\emph{self decay} of the  quantum fluctuations of the inflaton
during slow roll (quasi de Sitter) expansion. In this case,
$\Delta$ is a simple function of the slow roll parameters. For
superhorizon modes we find that the amplitude of the  inflaton
quantum fluctuations relaxes as a power law $\eta^{\Gamma}$ in
conformal time. To lowest order in slow roll, we find $ \Gamma $
completely determined by slow roll parameters and the amplitude of
the power spectrum of curvature perturbations $
\triangle^2_{\mathcal{R}} $: \be \Gamma = \frac{8 \; \xi^2_V \;
\triangle^2_{\mathcal{R}}}{(\epsilon_V-\eta_V)^2}
\left[1+\mathcal{O}(\epsilon_V,\eta_V)\right]
 \ee
\noindent where $\eta$ is conformal time and
$\xi_V,\eta_V,\epsilon_V$ are the standard slow roll parameters.
As a consequence, the growing mode which dominates at late time
evolves as 
\be 
\frac{\eta^{\eta_V-\epsilon_V +\Gamma}}{\eta} \;.
\ee 
featuring a {\it new scaling dimension} $\Gamma$ slowing down
the growth of the dominant mode.

 The decay of the inflaton quantum fluctuations with wavelengths
 deep within the Hubble radius during slow roll inflation is
 {\bf enhanced} by the infrared behavior associated with the collinear
 emission of ultrasoft  quanta, namely \emph{ bremsstrahlung
radiation of superhorizon fluctuations}. The decay  hastens as the
physical wavelength approaches the horizon because the phase space
for the emission of superhorizon quanta opens up as the wavelength
nears horizon crossing.}

\item{We discuss the implications of these results for scalar and
tensor perturbations, and establish a connection with previous
calculations of non-gaussian correlations.}
\end{itemize}

\section{Quantum Inflaton Decay}

We consider a general interacting scalar quantum field theory with cubic
couplings in a spatially flat cosmological Friedmann-Robertson-Walker
 space time with scale factor $a(t)$. The cubic
couplings are the lowest order non-linearities. Our study applies to
 two different scenarios, i) the inflaton  $\phi$ coupled to
 another scalar field $\varphi$, ii) the inflaton field self-coupled
 via a trilinear coupling. We consider the fields to be minimally
 coupled to gravity.

In comoving coordinates the action for case i) is given by
\bea\label{2fields}
&&A= \int d^3x \; dt \;  a^3(t) \Bigg\{ \frac{1}{2} \;
{\dot{\phi}^2}-\frac{(\nabla \phi)^2}{2a^2}-\frac{1}{2} \; M^2 \;
\phi^2 + \cr \cr 
&&+\frac{1}{2} \; {\dot{\varphi}^2}-\frac{(\nabla
\varphi)^2}{2a^2}-\frac{1}{2} \; m^2 \; \varphi^2 - g \;  \phi \,
\varphi^2 +J(t) \; \phi + \cr \cr 
&&+\mathrm{higher \; nonlinear \; terms}
\Bigg\}
\eea
\noindent  and for the case ii),
\bea\label{1field}
&&A= \int d^3x \;  dt \;  a^3(t) \Bigg\{ \frac{1}{2} \;
{\dot{\phi}^2}-\frac{(\nabla \phi)^2}{2a^2}-\frac{1}{2} \; M^2 \;
\phi^2 + \cr \cr
&&+\frac{g}{3} \;  \phi^3 +J(t) \; \phi+  \mathrm{higher \;
nonlinear \; terms} \Bigg\}
\eea
\noindent The linear term in $\phi$ is a counterterm that will be
used to cancel the tadpole diagram in the equations of motion. The
higher nonlinear terms do not affect our results but they are
necessary to stabilize the theory.

We computed in these two models the quantum decay of the inflaton
from the self-energy corrections to the equations of
motion to one-loop order\cite{nos}.

We worked in conformal time $\eta$ with $d\eta =
dt/a(t)$ and introduce a conformal rescaling of the fields
$$
a(t) \; \phi(\vx,t) = \chi(\vx,\eta) ~~;~~
a(t) \; \varphi(\vx,t)=\delta(\vx,\eta)  \; .
$$
The action Eq. (\ref{2fields}) (after discarding surface terms that
do not affect the equations of motion) reads:
\bea\label{confoaction}
&&A\Big[\chi,\delta\Big]= \int d^3x \;  d\eta \;
\Bigg\{ \frac{{\chi'}^2}{2}-\frac{(\nabla
\chi)^2}{2}\cr \cr
&& -\frac{\mathcal{M}^2_{\chi}(\eta)}{2} \; \chi^2 + 
\frac{{\delta'}^2}{2}-\frac{(\nabla
\delta)^2}{2}-\frac{\mathcal{M}^2_{\delta}(\eta)}{2} \; \delta^2 \cr \cr
&& -g \; C(\eta) \;  \chi \; \delta^2 + C^3(\eta) \;  J(\eta) \;  \chi\Bigg\}
\eea
\noindent with primes denoting derivatives with respect to conformal
time $\eta, \; C(\eta)= a(t(\eta))$ being the scale factor as a
function of $\eta$ and
\bea 
&&\mathcal{M}^2_{\chi}(\eta)  =  M^2 \;
C^2(\eta)-\frac{C''(\eta)}{C(\eta)} \; , \cr \cr 
&&\mathcal{M}^2_{\delta}(\eta)  =  m^2 \;
C^2(\eta)-\frac{C''(\eta)}{C(\eta)}  \; .\label{massdelta}
\eea
\noindent For
de Sitter space time, the scale factor is given by:
\be \label{scalefactor}
a(t)= e^{Ht} \; ,  \; C(\eta) = -\frac{1}{H\eta} \; , \; 
\eta= -\frac{e^{-Ht}}{H} \; ,
\ee
\noindent with $H$ the Hubble constant and where $\eta = -\frac{1}{H} $ 
corresponds to the initial time $t=0$.

The Heisenberg equations of motion for the Fourier field modes of
wave vector $k$ in the free ($g=0$) theory are given by
\bea
\chi''_{\vk}(\eta)+
\Big[k^2-\frac{1}{\eta^2}\Big(\nu^2-\frac{1}{4} \Big)
\Big]\chi_{\vk}(\eta)& = & 0 \; , \cr  \cr
\delta''_{\vk}(\eta)+
\Big[k^2-\frac{1}{\eta^2}\Big(\bnu^2-\frac{1}{4} \Big)
\Big]\delta_{\vk}(\eta)& = & 0 \; , \nonumber
\eea \noindent
where 
\be \label{nus} 
\nu^2  =  \frac{9}{4}- \frac{M^2}{H^2} \quad , \quad  
\bnu^2  = \frac{9}{4}-\frac{m^2}{H^2}  \; . 
\ee 
The Heisenberg free field operators can be expanded in terms of the
linearly independent solutions of the mode equation 
\be
S''_{\nu}(k;\eta)+ \Big[k^2-\frac{1}{\eta^2}\Big(\nu^2-\frac{1}{4}
\Big) \Big]S_{\nu}(k;\eta) =  0\, , \label{modeeqn} 
\ee \noindent
for $\nu,\bnu$ respectively. We choose the usual Bunch-Davies
initial conditions at $ \eta \to - \infty $ 
for the mode functions, namely the usual plane
waves for wavelengths deep inside the Hubble radius $|k \;
\eta|\gg 1$. The mode functions $S_{\nu}(q,\eta)$ associated with
the Bunch-Davies vacuum are given by 
\be\label{Snu}
S_{\nu}(k,\eta)= \frac{1}{2}\; i^{-\nu-\frac{1}{2}} \sqrt{\pi
\eta} \; H^{(2)}_\nu(k\eta) \; . 
\ee 
The modes with index $ \nu $ are associated to the inflaton field $ \phi $ 
while the  modes with index $ \bnu $ are associated 
to the lighter field $ \varphi $. In the case of the de Sitter background
eq.(\ref{scalefactor}), $ \nu $ and $ \bnu $ are given by eqs.(\ref{nus}).

We showed in ref.\cite{nos} that during slow-roll inflation the 
background  is quasi de Sitter and to lowest order in slow roll for the scalar 
perturbations is given by:
\be\label{quasiDS}
C(\eta)=-\frac{1}{H \; \eta} (1+\epsilon_V) + \mathcal{O}(\epsilon_V^2)
\ee
where $ \epsilon_V $ is the usual the slow roll parameter (see for example
\cite{clar}). Therefore, for model ii) the parameters $ \nu $ and 
$ \Delta $ associated to the inflaton are given by\cite{nos}
\bea\label{Dig}
\nu &=& \frac{3}{2}+\epsilon_V- \frac{M^2}{3 \, H^2} =
\frac{3}{2}+\epsilon_V-\eta_V \; , \cr \cr 
\Delta&\equiv& \frac32 - \nu=\eta_V -\epsilon_V \; ,
\eea
where $ \eta_V $ is the usual the slow roll parameter (see for example
\cite{clar}). Notice that the CMB anisotropy observations indicate that
the slow roll parameters are of the order $ 10^{-2} $ and hence 
$\Delta$ is of the order $  10^{-2} $ too . 

For gauge invariant scalar and tensor perturbations the 
infrared parameter $\Delta$ takes different forms. We find
to first order in slow roll\cite{nos}:
\bea
\Delta &=& -\eta_V -\epsilon_V \quad {\rm scalar~ gauge~invariant
~pert.}
\cr \cr
\Delta &=& -\epsilon_V \quad {\rm tensor~ 
gauge~invariant~pert.} \nonumber
\eea

In ref.\cite{nos} we obtained the decay law for the quantum fluctuations 
of the inflaton field $\phi$ by using the equation of motion for the 
expectation value of $\phi$ and implementing the dynamical renormalization group
(DRG). These equations follow from
the non-equilibrium generating functional which
 involves  forward and backward time evolution, typical
 of a density matrix. Unlike the S-matrix  case (which is an in-out
transition  probability where only forward time evolution is required),
the time evolution of an expectation value is an initial value problem which
requires an in-in matrix element. Real time
 equations of motion obtained from the non-equilibrium generating
 functional are guaranteed to be retarded. In Fourier space
the equations of motion for the expectation value of the field
$$
X_{\vk}(\eta) \equiv \langle \chi_{\vk}(\eta)\rangle \; ,
$$
are the integro-differential equation\cite{nos}  
\bea \label{ecmov}
X''_{\vk}(\eta)&+&\left[k^2-\frac{\nu^2_R-\frac{1}{4}
}{\eta^2}\right] X_{\vk}(\eta)+ g^2 \; \frac{\delta
M^2_1}{H^2\,\eta^2} \; X_{\vk}(\eta)  \cr \cr
&+&\frac{2 \, g^2 }{\eta\;H^2}\;
\int_{\eta_0}^{\eta} \frac{d\eta'}{\eta'} \;
\mathcal{K}_{\bnu}(k;\eta,\eta') \; X_{\vk}(\eta')=0\;, 
\eea
where $ \delta M^2_1 $ stands for the UV mass renormalization and
the kernel $ \mathcal{K}_{\bnu}(k;\eta,\eta') $ is given by the one-loop
self-energy diagram. The kernel $\mathcal{K}_{\bnu}(k;\eta,\eta')$ 
was computed in eq.(B14) of ref.\cite{nos}. 
It features a simple pole at $ {\bar\Delta} = 0$:
\bea
\mathcal{K}_{\bnu}(k;\eta,\eta') =
\frac{1}{4 \, \pi^2 \, k^3 \, (\eta \, \eta')^2 \;  {\bar\Delta}} 
\left\{ k(\eta-\eta') \cos[k(\eta-\eta')] \right. && \cr \cr
 -\left. (1 + k^2 \, \eta \, \eta')
\sin[k(\eta-\eta')] \right\} + {\cal O}( {\bar\Delta}^0) && \; .\nonumber
\eea
The perturbative solution of Eq.(\ref{ecmov}) is obtained by writing 
\be\label{pertsol} X_{\vk}(\eta)=
X_{0,\vk}(\eta)+g^2 \; X_{1,\vk}(\eta)+\mathcal{O}(g^4) 
\ee
The first order correction $ X_{1,\vk}(\eta) $ can be expressed by 
quadratures\cite{nos}
\bea\label{x1}
X_{1,\vk}(\eta)=-
\int_{\eta_0}^{0} d\eta' \; \mathcal{G}_{\nu}(k;\eta,\eta') \;
\left[ \frac{\delta M^2_1}{H^2\,\eta'^2} \; X_{0,\vk}(\eta')\right. &&\\ \cr
+\left. \frac{ 2
}{H^2\;\eta'}\int_{\eta_0}^{\eta'} \frac{d\eta''}{\eta''} \;
\mathcal{K}_{\bnu}(k;\eta',\eta'') \; X_{0,\vk}(\eta'')\right] &&\; . \nonumber
\eea
\noindent where $\mathcal{G}_{\nu}(k;\eta,\eta')$ is the retarded Green's function
$$
\mathcal{G}_{\nu}(k;\eta,\eta') = \theta(\eta-\eta') \; \frac{\pi}{2}
\; {\rm Im}\left[ H^{(1)}_{\nu}(k \, \eta) \;  
H^{(2)}_{\nu}(k \, \eta') \right] \; ,
$$
and $  H^{(1)}_{\nu}(z) $ and $ H^{(2)}_{\nu}(z) $ stand for
Hankel functions.

\section{Superhorizon Modes: $k=0$}

For superhorizon modes ($k=0$) the general solution of the unperturbed mode equations 
(\ref{modeeqn}) is given by
\bea \label{GF0}
X_{0,\vec{0}}(\eta) = A\;(-\eta)^{\beta_+}+B\;(-\eta)^{\beta_-} \; ;
\; \beta_{\pm} = \frac{1}{2}\pm \nu \; . \nonumber
\eea
We find for the first order correction $\mathcal{O}(g^2)$
from eqs.(\ref{pertsol}) and 
(\ref{x1})\cite{nos}
 \bea\label{solg2}
X_{\vec{0}}(\eta)=X_{0,\vec{0}}(\eta)\left[1+\Gamma \;
 \ln\frac{\eta}{\eta_0}+\mathrm{non-secular~terms}\right] &&\nonumber
\eea
\noindent with
\bea\label{Gamma_2}
&&\Gamma = \frac{g^2}{16\pi\,H^2\,\nu}\tan[\pi\,\nu] \left[1+
\frac{4}{\frac{9}{4}-\nu^2}\right] \cr \cr
&&= \frac{g^2}{16\pi\,H^2\,\nu}
\tan[\pi\,\nu]\left[1+ \frac{4\,H^2}{M^2}\right] ~~.
\eea
The term in $ \ln\eta $ is a secular term since it grows unbounded with time
implying a breakdown of perturbation theory. 
The {\it dynamical renormalization group resummation}\cite{DRG} exponentiates 
the secular terms in Eq.(\ref{solg2}) and leads to the improved solution\cite{nos},
\bea
&&X_{\vec{0}}(\eta)=
\left[\frac{\eta}{\eta_0}\right]^{\Gamma}\Bigg\{A(\eta_0)
\;  (-\eta)^{\beta_+} [1+\mathcal{O}(g^2)] \cr \cr
&& +B({\eta}_0) \;
(-\eta)^{\beta_-}[1+ \mathcal{O}(g^2)]\Bigg\} \label{resumX0ren}
\eea
The first term inside the square bracket in Eq. (\ref{Gamma_2})
(namely the unit term) corresponds to the case in which the
inflaton decays into massless particles conformally coupled to
gravity \cite{nos}.

The calculation leading to eq.(\ref{Gamma_2}) is valid for $ \bnu \to \frac32 $
(namely, $ m \ll H $) and we keep $ \nu $ as well as $M$ arbitrary.
We can analytically continue the formula (\ref{Gamma_2})
to $ H < M $ and then take the $ m \ll H \ll M $ limit. In this limit
$\Gamma$ becomes the decay rate of a particle with mass $M$ into massless
particles in Minkowski space-time:
$$
\lim_{m \ll H \to 0} H \; \Gamma = \Gamma_{Mink} =  g^2/(16 \pi M) \quad ,
$$
as it must be.

\section{Conclusions and further questions}

We have reviewed here particle decay of fields
minimally coupled to gravity in the case when the mass of the fields
is $\ll H$ during inflation. Unlike the  decay into  massless
fields conformally coupled to gravity, this case features a strong
infrared behavior which leads to novel results.

We have implemented in ref.\cite{nos} the dynamical renormalization group 
resummation program introduced in ref.\cite{DRG} combined with an expansion
in a small parameter $\Delta$ which regulates the infrared.

In the case of exact de Sitter inflation, $\Delta$ is a constant equal to
the ratio of the mass squared of the decay products to the Hubble 
constant squared,
while in slow roll inflation $\Delta$ is a simple function of slow
roll parameters. The expansion in $\Delta$ is akin to the
$\varepsilon$ expansion in critical phenomena in dimensional
regularization. The dynamical renormalization group provides a
resummation of the long-time secular divergences which determine the
decay law of quantum fluctuations.

The lack of energy conservation in an expanding
cosmology leads to the lack of kinematic
thresholds for particle decay. In particular, this possibility leads to the
\emph{self-decay} of quantum fluctuations whenever a
self-interaction is present.

We have studied the decay of a particle for a  cubic selfcoupled
scalar field in de Sitter space-time and applied the results to the
\emph{self-decay} of the inflaton quantum fluctuations
during quasi de Sitter, slow roll inflation.
We focused on extracting the decay law both for
wavelengths well inside and well outside the Hubble radius. In both
cases the strong infrared behavior enhances the decay.

The decay of fluctuations with wavelengths much smaller than the
Hubble radius {\bf is enhanced} by the collinear emission of
ultrasoft quanta, this process is identified as
 \emph{bremsstrahlung radiation} of superhorizon quanta. As the
 physical wavelength approaches the horizon, the phase space for
 this process opens up becoming the dominant decay channel for short
 wavelength modes in the region
 \be
H \ll k_{ph}(\eta) \lesssim \frac{H}{\eta_V-\epsilon_V} \; .
\ee
The decay of short wavelength modes hastens as the physical
wavelength approaches the horizon as a consequence of the opening up
of the phase space.

Superhorizon quantum fluctuations decay as a power law $\sim
\eta^{\Gamma}$ in conformal time, where $\Gamma$ is determined by
the following combination of the slow roll parameters and the
amplitude of curvature perturbations \be \Gamma = \frac{32 \;
\xi^2_V \; \triangle^2_{\mathcal{R}}}{(n_s-1+\frac{r}{4})^2}
\left[1+\mathcal{O}(\epsilon_V,\eta_V)\right]
 \ee
This decay law entails that the growing mode for superhorizon
wavelengths evolves as $  {\eta^{\eta_V-\epsilon_V +\Gamma}}/{\eta}
$ hence $\Gamma$ provides a {\it new scaling dimension}
slowing down the growing mode for late times $ \eta \to 0^- $.

The recent WMAP data indicate that 

$ 3. \times 10^{-8}  \gtrsim \Gamma \gtrsim 3.6 \times 10^{-9} $. 

This corresponds in cosmic time to a decay rate

$ 10^7 GeV \gtrsim \Gamma_{cosmic} \equiv H \; \Gamma \gtrsim 10^6 $GeV.

Although these values may seem small, it must be noticed that the
decay is a {\it secular}, namely  {\it cumulative} effect.

Forthcoming  observations of CMB anisotropies as well as large
scale surveys with ever greater precision will provide a
substantial body of high precision observational data which may
hint at corrections to  the generic and robust predictions of
inflation. If such is the case these observations will  pave the
way for a better determination of inflationary scenarios. Studying
the possible observational consequences of the quantum phenomena
reviewed here will therefore prove a worthwhile endeavor.

\end{document}